\documentclass[intlimits,twoside,a4paper]{article}

\usepackage[greek,english]{babel}
\usepackage{teubner}

\usepackage[eqsecnum]{cmpj3}

\usepackage{bm}

\issue{2024}{27}{3}{33604}
\doinumber{10.5488/CMP.27.33604}

\title[Revisiting the Lee-Yang singularities in the four-dimensional
  Ising model]%
      {Revisiting the Lee-Yang singularities in the four-dimensional
        Ising model: a tribute to the memory of Ralph Kenna}
\author[J. J. Ruiz-Lorenzo]{J. J. Ruiz-Lorenzo\orcid{0000-0003-0551-9891}} 
\address{
   Departamento de F\'{\i}sica and Instituto
  de Computaci\'on Cient\'{\i}fica
  Avanzada (ICCAEx), Universidad de Extremadura, 06006
  Badajoz, Spain}


\date{Received February 6, 2024, in final form May 10, 2024}

\begin{document}

\maketitle

\begin{abstract}

  We have studied numerically the Lee-Yang singularities of the four
  dimensional Ising model at criticality, which is believed to be in
  the same universality class as the $\phi_4^4$ scalar field
  theory. We have focused in the numerical characterization of the
  logarithmic corrections to the scaling of the zeros of the partition
  function and its cumulative probability distribution, finding a very
  good agreement with the predictions of the renormalization group
  computation on the $\phi_4^4$ scalar field theory. To obtain these
  results, we have extended a previous study [R. Kenna,  C. B. Lang,
    Nucl. Phys., 1993, \textbf{B393}, 461] in which there were computed
  numerically the first two zeros for $L\leqslant 24$ lattices, to the
  computation of the first four zeros for $L\leqslant 64$ lattices.
 

\printkeywords
\end{abstract}

\section{Introduction}

Since their introduction, the study of  complex singularities of
the free energy (equivalently the zeros of the partition function) in
magnetic field~\cite{yang:52,lee:52} and in
temperature~\cite{fisher:65,itzykson:89} has played a role of
paramount importance in understanding criticality in statistical
mechanics.

Regarding the complex singularities of the Ising model in a magnetic
field, we can recall some important results obtained by Yang and
Lee~\cite{yang:52,lee:52}.  They showed that the partition function of
the (ferromagnetic) Ising model defined on a regular graph of $N$
vertices (sites) in a complex magnetic field $H_{\text{c}}$ has its zeroes on
the unit circle of the fugacity variable ($z=\re^{\beta H_{\text{c}}}$), implying
a pure imaginary magnetic field $H_{\text{c}}=\ri H$, with real $H$. They also
found that in the paramagnetic phase these imaginary zeros condense at
a finite value of $H$ in the thermodynamic limit ($N \to \infty$),
known as the Yang-Lee edge. Furthermore, this edge reaches $H=0$ at
the critical point (when the Ising model is above its lower critical
dimension), and this is the origin of the singularity of the free
energy at criticality.

The use of these techniques, based on the study of  complex singularities,
can be cited both in numerical simulations
to characterize phase transitions in statistical mechanics and in quantum
field theory (e.g., see \cite{falcioni:82, marinari:84,
  marinari:98b}) and analytically, where, for example, they have been
employed to develop scaling relations in the presence of logarithmic
corrections~\cite{kenna:06a, kenna:06b,moueddene:24}.

Additionally, the emergence of these logarithmic corrections,
particularly in two-dimensional systems and models at their upper
critical dimensions~\cite{ruizlorenzo:98, kenna:06a,
  kenna:06b,ruizlorenzo:17}, has undergone thorough
investigation. This research has been facilitated by techniques
grounded in the properties of complex singularities.

An illustrative case of these investigations involved characterizing
the critical behavior of the four-dimensional Ising model. This model
is thought to belong to the same universality class as the $\phi^4_4$
lattice field theory~\cite{dominil:22}.  The significance of
$\phi^4_4$ in high-energy physics is paramount. For example, we can
cite its role in the triviality problem~\cite{callaway:88}.

Nevertheless, recent studies have raised questions about this
established scenario. In \cite{lundow:23}, the simulation of the
Ising model in both the canonical and microcanonical ensembles
suggests the possibility of the specific heat of the four-dimensional
Ising model being discontinuous, akin to mean-field behavior, in
contrast to its logarithmic divergence in the
$\phi^4_4$-theory. Furthermore, in \cite{akiyama:19}, employing
tensor renormalization group techniques, a weak first-order transition
scenario has been proposed\footnote{However, in
\cite{bittner:02}, the behavior of the susceptibility and the
specific heat, for $L \leqslant 40$, was confronted against the renormalization group
predictions, finding a very good agreement.}.

The primary objective of this paper is to numerically compute the
Lee-Yang (LY) singularities in the four-dimensional Ising model,
aiming to comprehensively understand the onset of logarithmic
corrections in the scaling of zeros at criticality as a function of  the
lattice size. In pursuit of this goal, we intend to expand upon the
initial work conducted by Kenna and Lang in \cite{kenna:93},
extending their research to larger lattices and incorporating the
computation of two additional zeros (the third and fourth ones).

Furthermore, given that we have computed the first four zeros, we can
analyze the cumulative probability distribution of the Lee-Yang
zeros and check its associated logarithmic correction.

The overarching objective is to assess whether the numerically
characterized logarithmic corrections align well with the analytical
predictions derived for $\phi^4_4$ using the renormalization group
(RG). This paper aims to provide support for the conventional
understanding that the Ising model and the $\phi^4$ field theory
belong to the same universality class in four dimensions.

The paper is organized as follows: in section~\ref{sec:model}, we provide
an overview of the model and observables. Following this, we present
relevant theoretical results in section~\ref{sec:teor}. Subsequently, in
section~\ref{sec:num}, we elaborate on the numerical simulations
conducted for this study. In section~\ref{sec:results}, we present our
findings, and the paper concludes with a section summarizing our
conclusions.

\section{The model and observables}
\label{sec:model}

We have considered the Ising model defined on a four-dimensional cubic
lattice with periodic boundary conditions, linear size $L$ and volume
$V=L^4$. The Hamiltonian of the model is given by
\begin{equation}\label{eq:H}
   {\cal H}=-\sum_{\langle\boldsymbol{x},\boldsymbol{y}\rangle} S_{\boldsymbol{x}} S_{\boldsymbol{y}}\,, 
\end{equation}
where $S_{\boldsymbol{x}}$ are Ising variables and the sum in
equation~\eqref{eq:H} runs over all pairs of lattice nearest-neighbors. As
usual, we denote the thermal average with $\langle (\cdots)\rangle$.

As stated in the Introduction, in order to compute the LY zeros we add a pure imaginary
magnetic field to the model, the new Hamiltonian being
\begin{equation}\label{eq:HH}
   {\cal H}_H=-\sum_{\langle\boldsymbol{x},\boldsymbol{y}\rangle} S_{\boldsymbol{x}} S_{\boldsymbol{y}}+ \ri H \sum_{\boldsymbol{x}}  S_{\boldsymbol{x}}\,, 
\end{equation}
with  $H\in \mathbb{R}$. For further use, we define the magnetization as
\begin{equation}
M=\sum_{\boldsymbol{x}}  S_{\boldsymbol{x}}\,.
\end{equation}
The partition function of the Hamiltonian ${\cal H}_H$ [equation (\ref{eq:HH})] is
\begin{equation}\label{eq:Z}
Z(\beta,H)=\sum_{[S_{\boldsymbol{x}}]} \exp\big(-\beta {\cal H} -\ri \beta H M\big)\,,
\end{equation}
that can be written as
\begin{equation}
Z(\beta,H) = Z(\beta,0)\big( \langle\cos(h M) \rangle - \ri\langle \sin(hM) \rangle \big)\,,
\end{equation}
with $h \equiv \beta H$ and the average $\langle (\cdots)\rangle$ is
computed in absence of the magnetic field (i.e., using the Hamiltonian
of equation (\ref{eq:H})). This fact implies $\langle \sin(hM) \rangle=0$
for all values of $h$ and that we will restrict our computation to the
case $h>0$.

Hence, the pure imaginary complex singularities in the magnetic field
(YL singularities) are determined by the solutions of the equation
\begin{equation}
\langle \cos(hM) \rangle=0\,.
\end{equation}
We will denote the $j$-th solution of this equation as $h_j$,
ordered in such a way that $h_{j+1}>h_j$. To solve
this equation, we have used the bisection method.

Once we have computed the LY zeros ($h_j$), the cumulative
distribution function of zeros can be easily computed as~\cite{janke:01}
\begin{equation}
G_L(h_j(L))=\frac{2 j -1}{2 L^d}\,,
\end{equation}
where $d$ is the dimensionality of the space ($d=4$ in this paper).
\section{Some theoretical results}
\label{sec:teor}
In this section we collect some analytical results obtained on the
$\phi^4_4$ field theory using RG, relevant to the scaling of the LY
zeros~\cite{kenna:06a,kenna:06b}.

The scaling behavior of the $j$-th LY zero with the reduced
temperature ($t=(T-T_c)/T_c$, $T_c$ being the critical temperature) is
\begin{equation}
h_j(t)\sim t^\Delta | \log t  |^{\hat \Delta}\,
\end{equation}
in the paramagnetic phase (i.e., for $t>0$).
The specific heat also behaves as
\begin{equation}
C_\infty(t) \sim t^{-\alpha} | \log|t| |^{\hat \alpha}\,.
\end{equation}

At the critical point the cumulative distribution function of the
zeros scales (in the thermodynamic limit) as~\cite{kenna:06a,kenna:06b}
\begin{equation}
  G_\infty(h)\sim h^{(2-\alpha)/\Delta}
  | \log h |^{\hat{\alpha}-(2-\alpha) \hat{\Delta}/\Delta}\,.
  \end{equation}
By inverting this equation, we obtain
\begin{equation}
  \label{eq:inv}
  h\sim G^{\Delta/(2-\alpha)} |\log G|^{-\theta} 
  \end{equation}
with  $-\theta \equiv \hat \Delta- \Delta {\hat \alpha}/(2-\alpha)$.

Since $G\sim L^{-d}$, the logarithmic correction of $h$
versus $L$ is the same as that of $h$ versus $G$, and 
the behavior at the infinite volume critical point of the LY
zeros in a finite box of size $L$ is
\begin{equation}
\label{eq:theta}
h_j(L) \sim  L^{-d \Delta/(2-\alpha)}  (\log L)^{-\theta}\,.
\end{equation}

The values of the reported exponents in the (one-component) $\phi_4^4$
theory are: $\alpha=0$, $\Delta=3/2$, ${\hat \alpha}=1/3$, ${\hat
  \Delta}=0$~\cite{zinn-justin:05,kenna:06a}. Hence, we can write the
following asymptotic relations for the four-dimensional Ising model
(which will be tested in the numerical part of this paper):
\begin{equation}
  \label{eq:scaling_hL}
h_j(L) \sim L^{-3} (\log L)^{-1/4}\,,
\end{equation}
and
\begin{equation}
    \label{eq:scaling_Gh}
G(h)\sim h^{4/3} |\log h|^{1/3} \,.
\end{equation}
Inverting the previous equation [or using equation (\ref{eq:inv})], we obtain
\begin{equation}
  \label{eq:scaling_hG}
h_j  \sim G^{3/4} |\log G |^{-1/4}\,.
\end{equation}

Since the main aim of this paper is to characterize the logarithmic
corrections of the LY zeros, we assume the RG values for $\Delta$
and $\alpha$ exponents and write equation (\ref{eq:scaling_hL}) as:
\begin{equation}
\label{eq:thetaL}
h_j(L) \sim  L^{-3}  (\log L)^{-\theta}\,,
\end{equation}
in this way we  try to evaluate the $\theta$ exponent analyzing
the numerical data and check if the data support the RG prediction:
$\theta=1/4$.

\section{Numerical simulations}
\label{sec:num}

We have investigated the model defined in equation~\eqref{eq:H} through
equilibrium numerical simulations. Specifically, we have brought to
thermal equilibrium our systems by combining cluster and local update
algorithms in order to  compute the LY singularities.

In particular, we have used a combination of the Wolff's single
cluster algorithm~\cite{wolff:89} with Metropolis
updates~\cite{sokal:97}. Our elementary Monte Carlo step on a lattice
of size $L$ is composed by $L^2$ Wolff's single-cluster updates and
subsequently by a full sweep Metropolis actualization of the lattice.

Furthermore, we have performed all our simulations at the infinite
volume critical inverse temperature $\beta_c=0.14969383(6)$~\cite{lundow:23}. Other previous estimates
were: $\beta_c=0.149694(2)$~\cite{stauffer:97,bittner:02} (based on
numerical simulations and analysis of high-temperature series
analysis) and that of Kenna and Lang
$\beta_c=0.149703(15)$~\cite{kenna:93} (based on the numerical study
of complex singularities)\footnote{In a recent
paper~\cite{akiyama:19} a very large four-dimensional Ising model,
$1024^4$, was numerically analyzed using the tensor renormalization
group reporting a critical inverse temperature of
$\beta_c=0.1503677(1)$ (statistically) very different from all the
previous published values.}.

To check the thermalization of our runs, we have monitored some non-local
observables as the second-moment correlation length and the susceptibility
as a function of the Monte Carlo time.
We have simulated a large number of runs on the same lattice
(different initial configurations and random numbers)\footnote{This strategy
allows us to optimize the use of clusters with a large number of
processors: the obvious drawback is the need of thermalizing the
different pseudosamples. However, the final outcome is positive.}.
In addition, we have computed the statistical error using the
jackknife procedure~\cite{young:15} merging the different
pseudosamples in ten groups.

Finally, details about the numerical simulations can be found in
table~\ref{tab:stat}.

\begin{table}[h]
	\caption{Parameters of the numerical simulations. $L$ is the lattice
		size, $N_{\mathrm{therm}}$ is the number of thermalizations
		(composed by $L^2$ Wolff updates complemented by a Metropolis
		sweep), $N_{\mathrm{steps}}$ is the number steps after the
		thermalization (in which we measure every 10 steps) and $N_{\mathrm
			{pseudo}}$ is the number of pseudo-samples (see the text). Hence,
		for a given lattice, we have performed $N_{\mathrm {pseudo}} \times
		N_{\mathrm{steps}}/10$ measures.}
	\vspace{2mm}
  \begin{center}
\begin{tabular}{c c c c}
  \hline   \hline
  $L$ & $N_{\mathrm{therm}}$ & $N_{\mathrm{steps}}
    $ & $N_{\mathrm {pseudo}}$ \\
    \hline
8  & 5120 & 10240 &   4240  \\
12 & 5120 & 10240 &   4010  \\
16 & 5120 & 10240 &   2540  \\
24 & 5120 & 10240 &   2470  \\
32 & 5120 & 10240 &   2230  \\
48 & 5120 & 10240 &   1720  \\
64 & 5120 & 10240 &   1780  \\
\hline   \hline
\end{tabular}
\label{tab:stat}
\end{center}
\end{table}

\section{Results}
\label{sec:results}
In the first two parts of this section we  analyze the scaling of the
four first zeros. In the third part we present the analysis of
their cumulative probability distribution.

\subsection{Scaling of the zeros}

\begin{figure}[h!]
  \begin{center}
 \includegraphics[width=0.6\columnwidth]{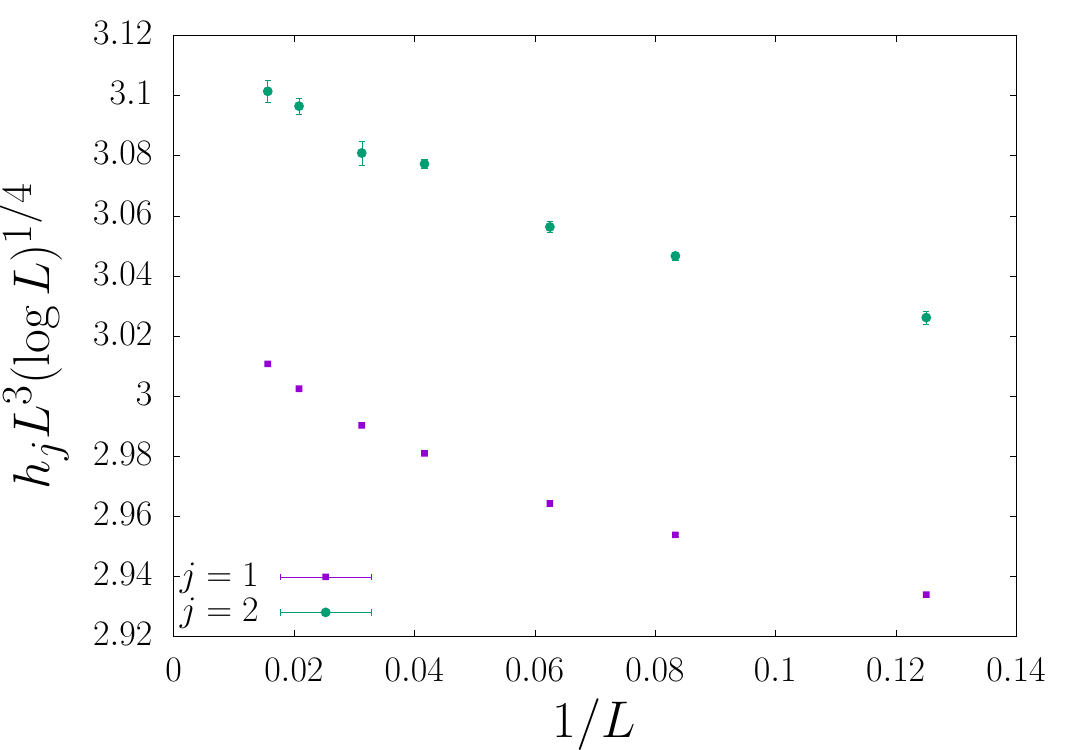}
\end{center}
 \caption{(Colour online) We show $h_j L^3 (\log L)^{1/4}$ versus $1/L$ in order to
   check the logarithmic correction [see equation (\ref{eq:scaling_hL})] for the $j$-th zero. If
   the data are in the asymptotic regime and the analytical prediction
   is correct, then $h_j L^3 (\log L)^{1/4}\simeq \mathrm{const}$. In
   this figure we plot only the behavior of the first two zeros. To
   have the data of these two zeros in the same figure, we have
   shifted vertically the $j=1$-points by an amount of 1.55. Notice
   that the tiny error bars allow us to see the subleading
   corrections (see the text).}
  \label{fig:scaling12}
\end{figure}

\begin{figure}[h!]
    \begin{center}
 \includegraphics[width=0.6\columnwidth]{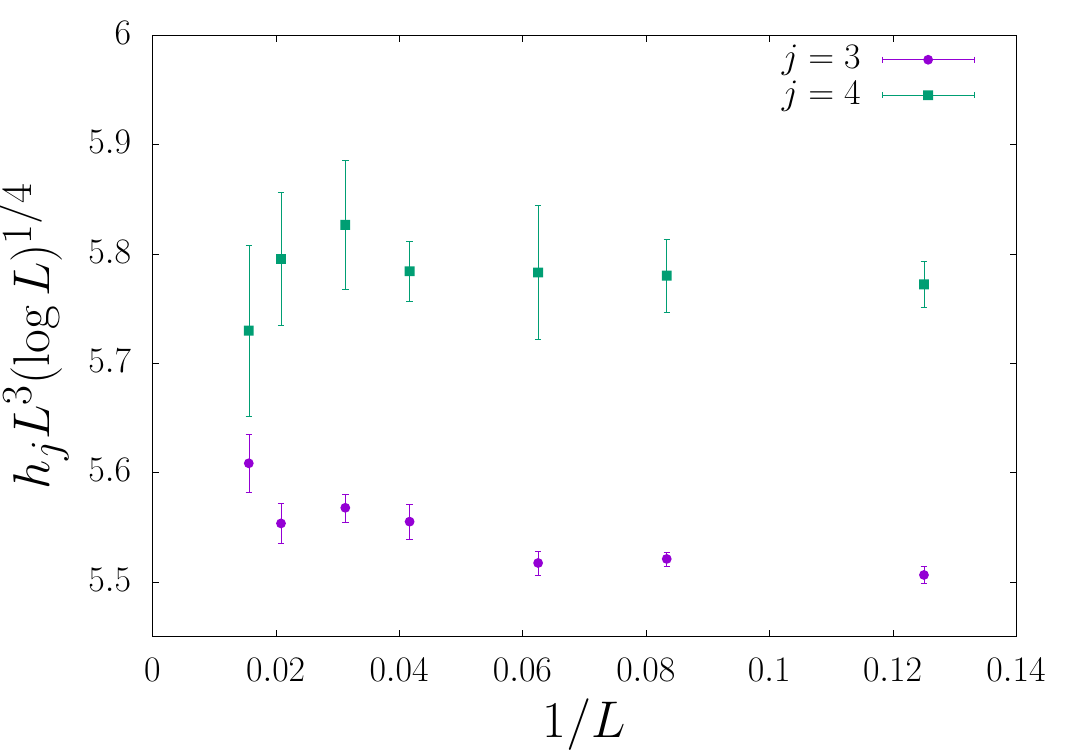}
\end{center}
 \caption{(Colour online) We show $h_j L^3 (\log L)^{1/4}$ versus $1/L$ in order to
   check the logarithmic correction [see equation (\ref{eq:scaling_hL})] for the $j$-th zero. In
   this figure we plot only the behavior of the third and fourth
   zeros. To have the data of these two zeros in the same figure, we have
   shifted vertically the $j=3$-points by an amount of 1.05.
   Notice that all the lattices follow the analytical
   prediction (no dependence with $1/L$) taking into account the error
   bars.}
  \label{fig:scaling34}
\end{figure}

In table \ref{table:zeros} we present the numerical values of the
first four LY zeros for the seven lattice sizes simulated\footnote{It
is not easy to compare numerically our values reported in table
\ref{table:zeros} with those of Kenna and Lang~\cite{kenna:93}. First,
we use a slightly different value of the inverse temperature. And
second, they used the spectral density method to find the zeros in
near values of the temperature which could bias the final
results. Overall, we found a good agreement with their computed
zeros.}. 
In figures \ref{fig:scaling12} and \ref{fig:scaling34} we show the
dependence with $L$ and the associated logarithmic corrections [see
equation (\ref{eq:scaling_hL})] by plotting $h_j L^3 (\log L)^{1/4}$ (that
should be constant) versus $1/L$.

\begin{table}[h]
	\caption{Values of the first four computed zeros for all the seven simulated lattice sizes.} 
\begin{center}
  \begin{tabular}{c c c c c}
    \hline   \hline
$L$ & zero \#1        & zero \#2       & zero \#3       & zero \#4     \\     \hline
8   & 0.0022510(6)    & 0.004922(3)    & 0.00725(1)     & 0.00939(4)   \\
12  & 0.0006470(2)    & 0.0014040(6)   & 0.002061(3)    & 0.00266(2)   \\
16  & 0.00026758(8)   & 0.0005782(3)   & 0.000845(2)    & 0.00109(1)   \\
24  & 0.00007753(4)   & 0.00016672(8)  & 0.0002441(9)   & 0.000313(2)  \\
32  & 0.000032214(9)  & 0.00006891(9)  & 0.0001010(3)   & 0.000130(1)  \\
48  & 0.000009363(5)  & 0.00001996(2)  & 0.0000290(1)   & 0.0000374(4) \\
64  & 0.000003902(2)  & 0.00000828(1)  & 0.00001218(7)  & 0.0000153(2) \\
\hline   \hline
  \end{tabular}
\end{center}
  \label{table:zeros}
\end{table}  

Only for the higher-order zeros ($j=3$ and $j=4$) we obtain $h_j L^3
(\log L)^{1/4}\simeq \mathrm{const}$ for all the lattice sizes
simulated (see figure \ref{fig:scaling34}) taking into account the
statistical errors of the data.
This initial analysis shows that the behavior of $h_j L^3 (\log
L)^{1/4}\simeq \mathrm{const}$ holds, hence, strongly supporting the
RG predictions. Notice that for the lower-order zeros, the points of
figure \ref{fig:scaling12} show a small variation of about 2--3\%.

\begin{figure}[h!]
  \begin{center}
    \includegraphics[width=0.6\columnwidth]{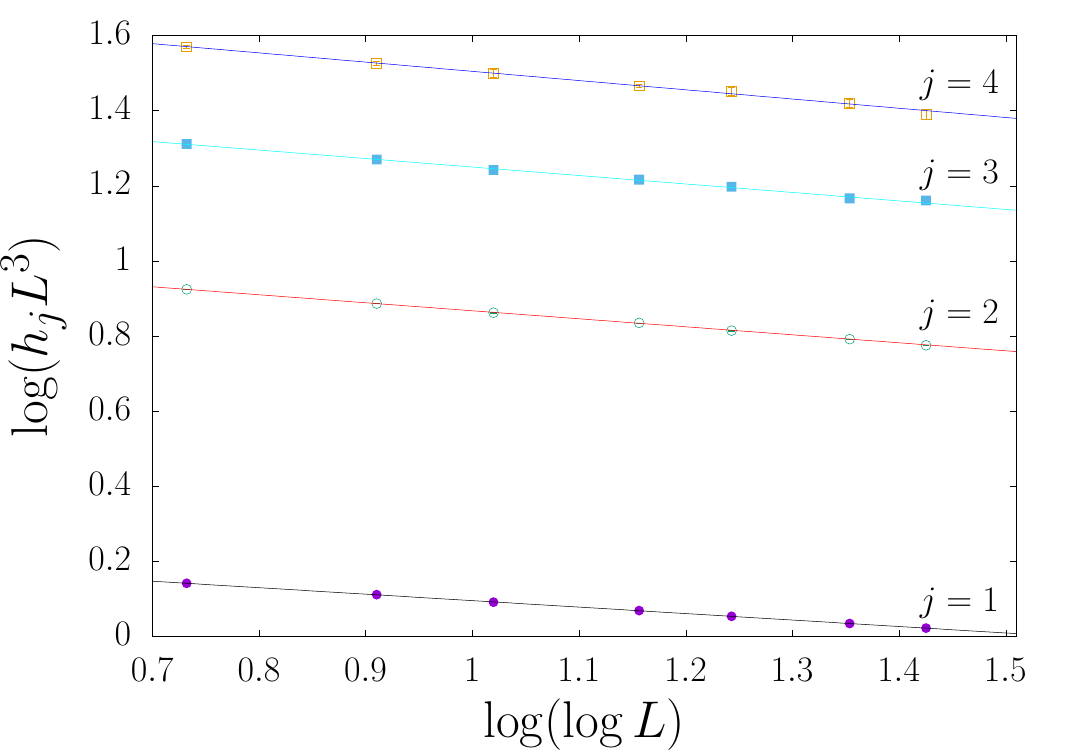}
  \end{center}
  \caption{(Colour online) $\log(h_j L^3)$ versus $\log \log L$ in order to compute
    the value of the $\theta$-exponent [see equation (\ref{eq:thetaL})] for
    the first four zeros.  The values of the different slopes
    are reported in table \ref{table:theta}. The size of the error
    bars is smaller than the ones of the symbols. }
  \label{fig:fits}
\end{figure}

As did by Kenna and Lang~\cite{kenna:93}, we can try to obtain the $\theta$
exponent [defined in equation (\ref{eq:thetaL})]. To do that, we plot in
figure \ref{fig:fits}, $\log(h_j L^3)$ versus $\log \log L$ to extract
the $\theta$ exponent via a linear fit.

The different slopes plotted in this figure are reported in table
\ref{table:theta}. Leaving free the $\theta$ exponent, we have
obtained good fits (see the first four columns of table
\ref{table:theta}). However, the value of the $\theta$ exponent is
clearly different from the RG prefiction for the lower-order zeros.
In addition, fixing the exponent $\theta$ to $1/4$ in the fit to
equation (\ref{eq:thetaL}) worsens the situation for the lower-order zeros (see also figure \ref{fig:scaling12}). 

The rationale for this behavior is as follows. We have been able to
compute with high precision the lower-order LY zeroes, while the
higher-order ones present, in comparison, greater relative
errors. This means that the possible subleading behaviors (for
example, corrections to scaling or, see below, improved scaling for
the leading term) are hidden inside their error bars. However, the
very small statistical errors of the lower-order zeros allow us to
start to see and analyze (see below) these additional behaviors.

In the next subsection we  try to improve this analysis.

\begin{table}[h!]
	\caption{Value of the $\theta$ exponent as a function of the order
		of the zero using equation (\ref{eq:thetaL}).  The first column is the
		order of the zero, the second column is the minimum value of the
		lattice size used in the fit, the third column reports the
		$\theta$ exponent obtained in the fit, and in the next column we
		report the goodness of the fit via the
		$\chi^2/\mathrm{d.o.f.}$ (where d.o.f. stands for number of degrees of freedom of the fit). In the last two columns, we report the
		minimum lattice size and the goodness of the fit
		$\chi^2/\mathrm{d.o.f.}$ by fitting the data to
		equation (\ref{eq:thetaL}) with $\theta$ fixed to its RG value
		($\theta=1/4$). For all the fits reported in this paper, we have
		computed the minimum value of $L$, in such a way the $p$-value of
		the least-squares fit is greater than $5\%$. Notice that for the
		first zero and fixed $\theta=1/4$, we have been
		unable to find an acceptable fit (i.e., we have obtained a  $p$-value $<5\%$).
	}
\vspace{2mm}
\begin{center}
  \begin{tabular}{c| c c c |c c}
    \hline \hline
 &\multicolumn{3}{c}{Free $\theta$} & \multicolumn{2}{c}{Fixed $\theta=1/4$} \\
    $j$-th zero & $L_\mathrm{min}$ & $\theta$ & $\chi^2/\mathrm{d.o.f.}$ & $L_\mathrm{min}$ & $\chi^2/\mathrm{d.o.f.}$ \\
  \hline   
1  &  8  &  0.1722(5)  & 8.4/5 & -  & -  \\
2  &  8  & 0.213(1)    & 7.0/5 & 48 & 1.18/1\\
3  &  8  & 0.226(5)   & 5.5/5 & 24  & 1.14/3\\
4  &  8  & 0.246(11)  & 1.1/5 &  8  & 0.21/6\\
\hline   \hline
  \end{tabular}
\end{center}
  \label{table:theta}
\end{table}

\subsection{Two improved scalings of the zeros}

In the previous subsection, to analyze the scaling of the zeros, we
have used the asymptotic scaling form given by
equation (\ref{eq:scaling_hL}).
In this subsection we re-analyze the data using two kinds of
improvements. The first one is to write the scaling relation 
improving the dependence on the logarithm (as suggested by the
theory), and the second one is based in the introduction of the leading
correction to scaling in the fits.

The starting point in our first approach is to note that the
integration of the $\beta$-function\footnote{$\beta(u)\equiv \rd u/\rd
\log b$. For the four-dimensional Ising model, the leading order of
this function is $\beta=-B u^2$, with $B>0$~\cite{luijten:97}.} for
the coupling $u$ in the $\phi_4^4$-model will provide a factor $(1+A
\log b)$ instead of its asymptotic form given by $\log b$ for large
$b$ ($b$ being the scaling factor in the RG
approach)~\cite{luijten:97}: $1/u(b) \propto 1+ A \log b$.  Using this
fact, we can rewrite equation (\ref{eq:scaling_hL}) as
\begin{equation}
\label{eq:thetaI}
h_j(L) \sim  L^{-3}  (1+ A\log L)^{-\theta}\,,
\end{equation}
for a suitable constant $A$ that should be independent of the order of
the zero\footnote{We thank an anonymous referee for pointing out to us
this improved scaling relation.}.

\begin{figure}[!t]
  \begin{center}
    \includegraphics[width=0.6\columnwidth]{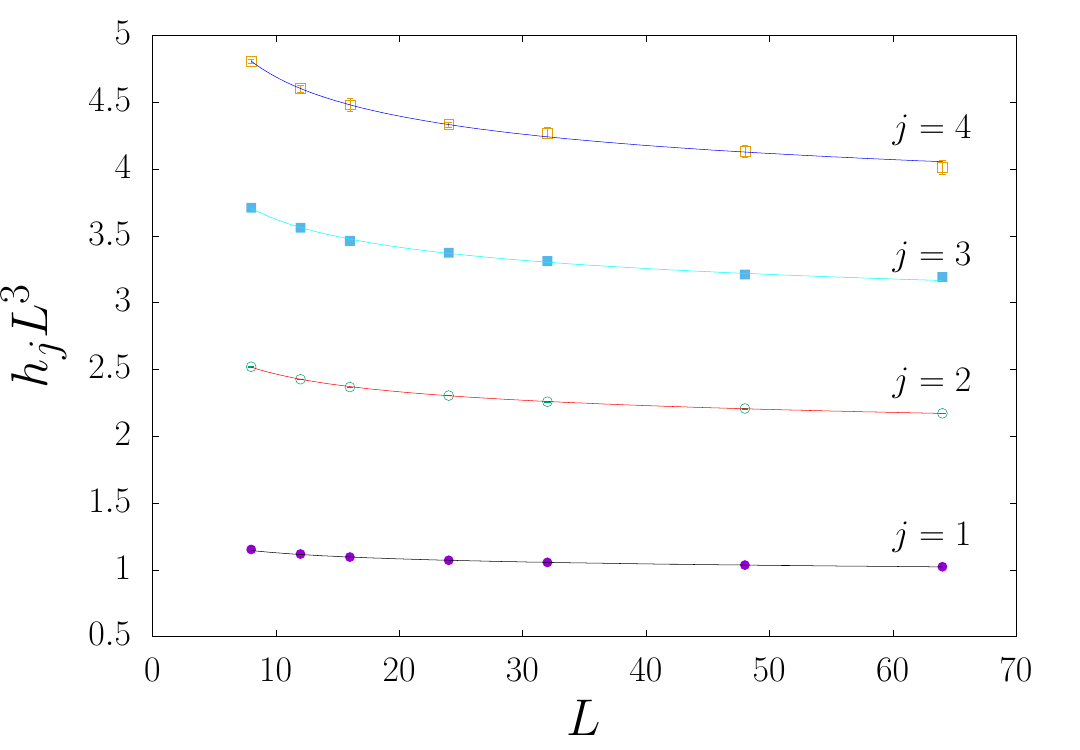}
  \end{center}
  \caption{(Colour online) We show $h_j L^3$ versus $L$ in order to check the
    improved scaling relation [see equation (\ref{eq:thetaI})] for the
    first four zeros. The continuous lines are fits to
    equation (\ref{eq:thetaI}) with $\theta=1/4$ (see table
    \ref{table:thetaI}). The size of the error bars is smaller than
    the ones of the symbols.}
  \label{fig:fitsI}
\end{figure}

In figure \ref{fig:fitsI} we present the behavior of $h_j L^3$ versus
$L$ to check this improved scaling behavior and the fits to
equation (\ref{eq:thetaI}) with $\theta$ fixed to its RG value of 1/4
(i.e., the improved leading prediction of the RG).  In addition, in
table \ref{table:thetaI} we report (see the last two columns) the
statistical quality of these four fits and the small lattice size for
which we have obtained  acceptable statistical fits. Notice that the four
zeros behave accordingly as stated by the improved RG prediction
(even the lower ones). Hence, this improved scaling is capable of capturing
the behavior even for the lower-order zeros despite their small
statistical errors.

However, we have found that the value of $A$ is roughly independent of
the order of the zero ($A=2.0(1)$ for $j=2$ and $3.5(7)$ for $j=3$,
unfortunately, for $j=4$ the relative error for $A$ is greater than 100\%) but not for
the first zero where $A= 0.64(1)$. The explanation of this fact is
that we would need to take into account the leading scaling
correction of the model (see below).

Finally, we have tried to obtain the value of the $\theta$ exponent
using this improved scaling. We present the results of these fits in
the 2nd to the 4th columns of table \ref{table:thetaI}. The obtained
$\theta$ exponents are compatible with the RG prediction. However,
being a three parameter fit, the statistical errors associated to the
different estimates of $\theta$ are large.

\begin{table}[h!]
	\caption{Value of the $\theta$ exponent as a function of the order
		of the zero using the improved scaling relation given by
		equation (\ref{eq:thetaI}).  The first column is the order of the zero,
		the second column is the minimum value of the lattice size used in
		the fit, the third column reports the $\theta$ exponent obtained
		in the fit, and in the next column we report the goodness of the
		fit via the $\chi^2/\mathrm{d.o.f.}$ In the last two columns, we
		report the minimum lattice size and the goodness of the fit
		$\chi^2/\mathrm{d.o.f.}$ by fitting the data to
		equation (\ref{eq:thetaI}) with $\theta$ fixed to its RG value
		($\theta=1/4$).}
	\vspace{2mm}
\begin{center}
  \begin{tabular}{c| c c c |c c}
    \hline \hline
 &\multicolumn{3}{c}{Free $\theta$} & \multicolumn{2}{c}{Fixed $\theta=1/4$} \\
    $j$-th zero & $L_\mathrm{min}$ & $\theta$ & $\chi^2/\mathrm{d.o.f.}$ & $L_\mathrm{min}$ & $\chi^2/\mathrm{d.o.f.}$ \\
  \hline   
1  &  8  & 0.174(5)   & 8.9/4 & 16 & 3.5/3\\
2  &  8  & 0.22(1)   & 6.8/4  & 8  & 10.2/5\\
3  &  8  & 0.23(5)   & 5.5/4  & 8  & 6.7/5\\
4  &  8  & 0.24(5)  & 1.1/4  & 8  &1.1/5\\
\hline   \hline
  \end{tabular}
\end{center}
  \label{table:thetaI}
\end{table}  

Now, we analyze the data using the second method of improvement
based in considering the effect of the leading scaling correction (at the
infinite volume critical point) which induces a correction term
proportional to\footnote{The
susceptibility can be computed using the behavior of the lowest zero
(see, for example, \cite{moueddene:24}): $\chi\sim L^{-d}
h_{j=1}(L)^{-2}$. Hence, the corrections to scaling of the LY zeros are
the same as those of the susceptibility. The analysis of the leading correction
to scaling for the susceptibility at the infinite volume critical
point can be found, for example,  in  equations (4.24) and (6.28) of~\cite{luijten:97} on the four-dimensional Ising model or in equation (39) of~\cite{ballesteros:98} on the four dimensional diluted Ising
model adapted to the pure case. Notice that if one considers the next leading
order contribution, $O(u^3)$, to 
the $\beta$-function, an additional  correction term 
$O(\log \log L/\log L)$ appears, this behavior was obtained by
R. Kenna in 2004~\cite{kenna:04}.} $O(u^{1/2})\propto 1/\sqrt{\log L}$.
Hence, the scaling of the $j$-th zero can be written as
\begin{equation}
  \label{eq:thetaC}
h_j(L) \sim  L^{-3} (\log L)^{-\theta}\left(1+\frac{C}{\sqrt{\log L}}\right)\,.
\end{equation}

In table \ref{table:thetaC} we report the results of our fits. Firstly,
we have tried to perform the fits to equation (\ref{eq:thetaC}) with a free
value of $\theta$ (see the 2nd, 3rd and 4th columns of the table
\ref{table:thetaC}).  Notice that the best fits for the lower-order
zeros provide lower values of the $\theta$ exponent, as in the
previous approaches.

In addition, we have fitted the data against equation (\ref{eq:thetaC}) with
$\theta=1/4$ (the RG prediction with the leading correction to
scaling term). The results of these fits can be seen in columns 5th
and 6th of table~\ref{table:thetaC}. Notice that the dependence on the
lattice size of all the zeros can be
described by the RG prediction, including the leading correction
to the scaling term.

\begin{figure}[!t]
  \begin{center}
    \includegraphics[width=0.6\columnwidth]{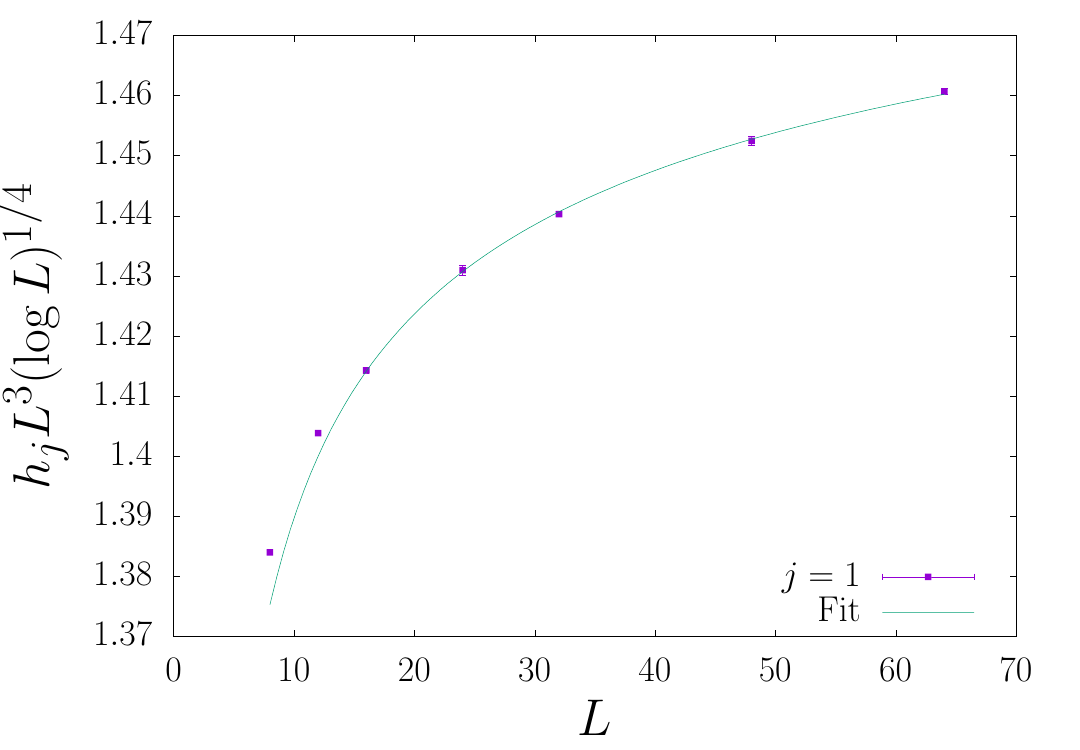}
  \end{center}
  \caption{(Colour online) We show $h_j L^3 (\log L)^{1/4}$ versus $L$ for the first
    zero in order to check the correction to scaling relation 
[see equation (\ref{eq:thetaC})]. The
    continuous line is the fit to equation (\ref{eq:thetaC}) with
    $\theta=1/4$ (see table \ref{table:thetaC}).}
  \label{fig:fitsC}
\end{figure}

To illustrate this point, we plot this fit for the first zero (the
most problematic case) in figure \ref{fig:fitsC}. The scaling is very
good for the  first zero taking $L\geqslant 16$, and we remark that the data
for this zero support the RG prediction.

To sum up, both improved methods produce similar (good) effects on the
different fits fixing $\theta=1/4$ (see the last two columns of 
tables \ref{table:theta}, \ref{table:thetaI} and \ref{table:thetaC}).
Obviously, both must be present in the analysis.
However, the fits using both improvements in the same equation yield
fitting constants with higher errors.
\newpage

\begin{table}[h!]
	\caption{Value of the $\theta$ exponent as a function of the order
		of the zero using the improved scaling relation given by
		equation (\ref{eq:thetaC}).  The first column is the order of the zero,
		the second column is the minimum value of the lattice size used in
		the fit, the third column reports the $\theta$ exponent obtained
		in the fit, and in the next column we report the goodness of the
		fit via the $\chi^2/\mathrm{d.o.f.}$ In the last two columns, we
		report the minimum lattice size and the goodness of the fit
		$\chi^2/\mathrm{d.o.f.}$ by fitting the data to
		equation (\ref{eq:thetaC}) with $\theta$ fixed to its RG value
		($\theta=1/4$).}
	\vspace{2mm}
\begin{center}
  \begin{tabular}{c| c c c |c c}
    \hline \hline
 &\multicolumn{3}{c}{Free $\theta$} & \multicolumn{2}{c}{Fixed $\theta=1/4$} \\
    $j$-th zero & $L_\mathrm{min}$ & $\theta$ & $\chi^2/\mathrm{d.o.f.}$ & $L_\mathrm{min}$ & $\chi^2/\mathrm{d.o.f.}$ \\
  \hline   
1  &  8  & 0.17(1)   & 8.1/4 & 16 & 2.1/3\\
2  &  8  & 0.23(2)   & 6.6/4  & 8  & 7.5/5\\
3  &  8  & 0.12(15)   & 4.4/4  & 8  & 6.1/5\\
4  &  8  & 0.32(19)  & 0.96/4  & 8  & 1.1/5\\
\hline   \hline
  \end{tabular}
\end{center}
  \label{table:thetaC}
\end{table}

\subsection{Density of the zeros}
In this subsection, we check the RG prediction (with no
improvements) in a different way, by introducing a $\lambda$ parameter
defined as\footnote{Notice that we test equation (\ref{eq:scaling_hG})
instead of equation (\ref{eq:scaling_Gh}). The reason is that it is simpler to
fit the data points $(x,y)$ with the statistical in the $y$-variables. In
this case, we have computed the statistical error on the $h_j$
variables, so we use them as $y$-variables in the fit and consequently
we test equation (\ref{eq:scaling_hG}).}
\begin{equation}
  h\sim \bigl[G^{3/4} (\log(G))^{-1/4}\bigl]^\lambda \,.
\end{equation}
If the prediction of RG holds, then $\lambda=1$.

\begin{figure}[h]
  \begin{center}
  \includegraphics[width=0.55\columnwidth]{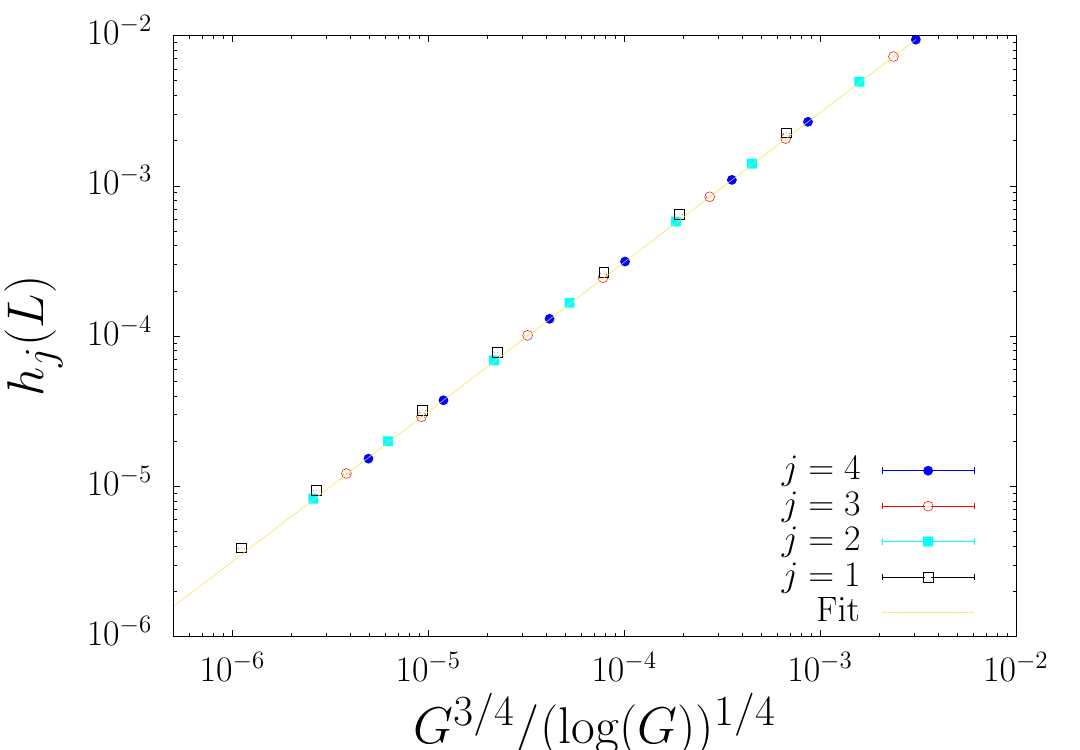}
\end{center}
  \caption{(Colour online) $h_j$ (for the first four zeros and the seven
    lattice sizes simulated) versus the analytical prediction
    $G^{3/4}/(\log(G))^{1/4}$, $G$ is the cumulative probability
    distribution of the zeros.  The fit has been computed (continuous line)
    using only  the
    fourth-zero data in a double-log scale. The slope of the straight
    line is the $\lambda$-parameter. We report in table
    \ref{table:lambda} the values of the $\lambda$-parameter as a
    function of the order of the zero.  The size of the error
    bars is smaller than the ones of the symbols.}
  \label{fig:Density}
\end{figure}

In figure \ref{fig:Density} we plot $h_j$ against
$G^{3/4}/(\log(G))^{1/4}$, where we have computed $G$ using the first
four zeros and the seven lattice sizes. The scaling of all the zeros
and sizes is very good,  all the points falling in a common curve.

\begin{table}[h]
	 \caption{The $\lambda$ parameter as a function of the order
		of the zero. The meaning of the different columns is the same as in
		table \ref{table:theta}.}
	\vspace{2mm}
\begin{center}
  \begin{tabular}{c c c c}
    \hline   \hline
    $j$-th zero & $L_\mathrm{min}$ & $\lambda$ & $\chi^2/\mathrm{d.o.f.}$ \\
  \hline
1  & 24 &  0.9943(2)   & 0.33/2\\
2  & 24 & 0.9966(4)    & 1.2/4 \\
3  &  8 & 0.9947(5)   & 4.7/5 \\
4  &  8 & 0.996(1)    & 2.0/5 \\
\hline   \hline
  \end{tabular}
\end{center}
  \label{table:lambda}
\end{table}  

We have performed a quantitative analysis of the slope (in a double
logarithmic scale) in order to compute $\lambda$. The results are
reported in table \ref{table:lambda}.

Notice that to avoid the use of the full covariance matrix in the
$\chi^2$ minimization fits (the data of the zeros belonging to the
same lattice size are correlated), we have performed independent fits
for all the four zeros (which are uncorrelated), obtaining four
estimates of the $\lambda$ parameter.
Overall, we find that the values of $\lambda$ point to $\lambda=1$,
accordingly with the RG prediction.

Finally, we remark that if the numerical data behave as $G(h)\sim a_1
h^{a_2}$ with $a_2 \simeq 1$ (for small $h$), we will be in presence of
a first order phase transition with a discontinuity in the
magnetization given by $a_1$. However, our numerical data are
compatible with $a_2=4/3$ (modified by logarithmic corrections with a
positive power)~\cite{janke:01} and we are unable to  see a trend towards a
first order transition in the analysis in the cumulative probability
distribution data.

\section{Conclusions}

By numerically computing the first four LY zeros of the
four-dimensional Ising model, we have been able to extend previous
numerical results but also  to unveil the onset of the
appearance of the logarithmic corrections associated with these
observables.

Specifically, the analysis of the scaling behavior of these zeros with
the lattice size indicates that the logarithmic corrections, as
foreseen by the renormalization group approach in a $\phi_4^4$
continuous field theory, manifest distinctly within the simulated
lattice sizes (where $L\leqslant 64$). In particular, the scaling is much
better using the improved versions of the scaling of the LY zeros (changing $\log L$ by $1+A \log L$  and the introduction of the leading scaling correction) and fixing $\theta=1/4$.

Another way to study the scaling of the LY zeros is to analyze the
properties of the cumulative probability distribution. Our findings
demonstrate once again a robust alignment with the predictions of the
field theory. Notably, as in the $h_j(L)$ analysis, we observed that
even for relatively modest lattice sizes, the zeros exhibit the
expected logarithmic power dependencies.   We remark that
the analyses of $h_j(L)$ and $h_j(G)$ are not independent.

In particular, the behavior of $h_j(G)$ is inconsistent with a second
order phase transition with a specific heat discontinuity at the
transition point ($\alpha=\hat \alpha=0$) and also with a weak first
order phase transition.  Notice that we have tested the following
combination of critical exponents (just in the scaling of $h_j$ versus
$L$ or in the scaling of $h_j$ versus $G$) $\hat \Delta- \Delta {\hat
  \alpha}/(2-\alpha)$ which mixes the odd ($\Delta$, $\hat \Delta$) and
even ($\alpha$, $\hat \alpha$) sectors of the theory, and this
combination of critical exponents is sensitive to the behavior of the
specific heat (exponents $\alpha$ and $\hat \alpha$).
Furthermore, we would like to remark that we have been unable to
detect a departure of the scaling of the zeros (as a function of the
lattice size) from the RG predictions.

Finally, the results presented in this paper are fully compatible with a second
order phase transition for the four-dimensional Ising model with the
exponents and logarithmic exponents predicted by the RG analysis of
the $\phi_4^4$ continuous field theory.

\section*{Acknowledgements}

I dedicate this paper to the memory of my friend and colleague
Ralph Kenna. I will always remember his vibrant energy, creative
imagination, technical prowers, and wide-ranging curiosity across
various fields of knowledge.

I would also like to express my heartfelt support to Claire and
Ro\'{i}s\'{i}n.

Our simulations have been carried out at the the \textit{Instituto de
  Computación Científica Avanzada de Extremadura} (ICCAEx), at
Badajoz. We would like to thank its staff.

This work was partially supported by Ministerio de Ciencia,
Innovaci\'on y Universidades (Spain), Agencia Estatal de
Investigaci\'on (AEI, Spain, 10.13039/501100011033), and European
Regional Development Fund (ERDF, A way of making Europe) through Grant
PID2020-112936GB-I00 and  by the Junta de Extremadura (Spain) and Fondo
Europeo de Desarrollo Regional (FEDER, EU) through Grant No.\ IB20079.

%

\ukrainianpart

\title{Перегляд сингулярностей Лі-Янга у чотиривимірній 
	моделі Ізінга: до вшанування пам'яті Ральфа Кенни}
\author{Х. Х. Руїс-Лоренцо}
\address{Фізичний факультет та Інститут перспективних наукових досліджень, Університет Естремадури, 06006, Бадахос, Іспанія}

\makeukrtitle

\begin{abstract}
	\tolerance=3000%
	Ми чисельним чином дослідили сингулярності Лі-Янга чотиривимірної
	моделі Ізінга в околі критичності, яка, як вважається, належить до
	того самого класу універсальності, що й скалярна теорія поля $\phi_4^4$. Ми зосередилися на числовій характеристиці логарифмічних поправок до масштабування нулів статистичної суми та її кумулятивного розподілу ймовірностей, отримавши дуже добре узгодження з передбаченнями теорії ренормгрупи в рамках
	$\phi_4^4$ теорії скалярного поля. Щоб отримати ці результати, ми узагальнили попереднє дослідження [R. Kenna, C. B. Lang, Nucl. Phys., 1993, \textbf{B393}, 461], в якому було обчислено перші два нулі для ґраток $L\leqslant 24$, аж до розрахунку перших чотирьох нулів для ґраток $L\leqslant 64$.
	\keywords ренормгрупа, скейлінг, логарифми, середнє поле, комплексні сингулярності
	
\end{abstract}

\end{document}